\documentclass[aps,showkeys]{revtex4}

\usepackage{epsfig}
\usepackage{eepic}
\usepackage{latexsym}
\usepackage{rotating}
\usepackage{lscape}
\usepackage{multirow}
\usepackage{subfigure} 
\usepackage{graphics}
\usepackage{dcolumn}
\usepackage{bm}
\newcommand{\beq}{\begin{equation}}
\newcommand{\eeq}{\end{equation}}
\newcommand{\bd}{\begin{displaymath}}
\newcommand{\ed}{\end{displaymath}}

\newcommand{\bei}{\begin{itemize}}
\newcommand{\eei}{\end{itemize}}

\newcommand{\bee}{\begin{enumerate}}
\newcommand{\eee}{\end{enumerate}}

\begin{document}

\noindent
\title{A study of the irradiation technique used for the external beam radiotherapy of retinoblastoma}

\author{L. Brualla$^1$, P.A. Mayorga$^2$, A. Fl\"uhs$^1$, A.M. Lallena$^3$, J. Sempau$^4$, W. Sauerwein$^1$\\
{\small {\it 
$^1$NCTeam, Strahlenklinik, Universit\"atsklinikum Essen, Hufelandstra\ss e 55, D-45122 Essen, Germany.\\
$^2$Departamento de F\'{\i}sica, Facultad de Ciencias, Pontificia Universidad Javeriana, Cra.\ 7 \#\ 40-62, Bogot\'a D.C., Colombia.\\
$^3$Departamento de F\'{\i}sica At\'omica, Molecular y Nuclear, Universidad de Granada, E-18071 Granada, Spain.\\
$^4$Institut de T\`ecniques Energ\`etiques. Universitat Polit\`ecnica de Catalunya, Diagonal 647, E-08028 Barcelona, Spain.
}}}

\date{\today}

\bigskip

\begin{abstract}
	
    {\bf Purpose:} Retinoblastoma is the most common eye tumor in childhood. The best outcome regarding tumor control and visual function can be reached by external beam radiotherapy. The benefits of the treatment are, however, jeopardized by a high incidence of radiation induced secondary malignancies and the fact that irradiated bones grow asymmetrically. In order to better exploit the advantages of external beam radiotherapy it is necessary to improve current techniques by reducing the irradiated volume and minimizing the dose to the facial bones. The purpose of this work is to evaluate the adequacy of Monte Carlo simulations and the accuracy of a commercial treatment planning system by means of experimental measurements. To this end dose measurements in water were performed using the dedicated collimator employed in the retinoblastoma treatment. Evaluation of these two calculation methods is necessary to further improve the irradiation technique. 

	{\bf Methods:} A Varian Clinac 2100 C/D operating at 6 MV is used. A dedicated collimator for the retinoblastoma treatment is inserted in the accessory tray holder. This collimator conforms a `D'--shaped off-axis field whose irradiated area can be either 5.2 or 3.1~cm$^2$. Depth dose distributions and lateral profiles were experimentally measured. Experimental results were compared with Monte Carlo simulations run with the {\sc penelope} code and with calculations performed with the analytical anisotropic algorithm implemented in the Eclipse treatment planning system using the gamma test.
	
	{\bf Results:} {\sc penelope} simulations agree reasonably well with the experimental data with discrepancies in the dose profiles less than 0.3~cm of distance-to-agreement and 3\% of dose. Discrepancies between the results found with the analytical anisotropic algorithm and the experimental data reach 0.3~cm and 6\%. The agreement in the penumbra region between the analytical anisotropic algorithm and the experiment is noticeably worse than the agreement between the latter and {\sc penelope}. The percentage of voxels with a gamma index larger than unity when comparing {\sc penelope} results with the experiment is on average 1\% assuming a 0.3~cm distance-to-agreement and a discrepancy of 3\% of dose. Under the same conditions the percentage of voxels exceeding a gamma index equal to one when the analytical anisotropic algorithm is considered is on average 7\%.
	
	{\bf Conclusions:} Although the discrepancies between the results obtained with the analytical anisotropic algorithm and the experimental data are noticeable, it is possible to consider this algorithm for routine treatment planning of retinoblastoma patients, provided the limitations of the algorithm are known and taken into account by the medical physicist. Monte Carlo simulation is essential for knowing these limitations. Monte Carlo simulation is required for optimizing the treatment technique and the dedicated collimator.
	
\end{abstract}


\maketitle

\section{Introduction}

Retinoblastoma is the most common eye tumor in childhood. Multiple, bilateral tumors are caused by a germline mutation at the RB-1 locus on chromosome 13 and therefore are hereditary~\cite{Friend1986}. Furthermore, this mutation leads also to secondary malignancies outside the eye. The best outcome for the primary tumors regarding tumor control and visual function can be reached by external beam radiotherapy~\cite{Sauerwein1997}. However, due to the specific genetic situation, a very high incidence (20\% within 30 years) of secondary malignancies after external beam radiotherapy has been observed~\cite{Abramson1998,Dommering2011,Turaka2011,Vasudevan2010}. Consequently, in the last two decades, major efforts were made to avoid external beam radiotherapy in these young children using especially systemic~\cite{Shields1996,Gallie1996} and recently even locally applied chemotherapy~\cite{Gobin2011} in combination with local treatments such as laser- and cryo-coagulation, and brachytherapy~\cite{Schueler2006a,Schueler2006b,Munier2008}. These techniques severely damage the retina, and therefore, have a negative influence on the visual acuity. Moreover, the long-term toxicity of chemotherapy on other organs cannot be neglected~\cite{Qaddoumi2012} and an increased incidence of leukemia and solid malignancies after chemotherapy has been reported~\cite{Gombos2007,Felix2001}. It seems therefore justified to reconsider specific external beam radiotherapy techniques aimed at reducing the irradiated volume which are designed to avoid asymmetric growth of the bones of the face. Schipper made in 1983 a first attempt in this direction by developing a highly precise lens sparing irradiation technique for retinoblastoma~\cite{Schipper1983,Schipper1997}. This technique, with some minor improvements, is still the state-of-the-art for the irradiation of retinoblastoma in children~\cite{Sauerwein2009}.

In this work an approach based on Monte Carlo (MC) simulation is introduced in order to assess the dose distribution of the mentioned radiotherapy technique. As a first step, the accuracy of the MC simulation is evaluated by means of experimental measurements in a water phantom. Secondly, the limitations of using the analytical anisotropic algorithm (AAA)~\cite{AAA,Ulmer1995,Ulmer2003} implemented in the Eclipse treatment planning system are evaluated by comparison with experimental data and MC results. The intention in the near future is to identify a linac-based gentle dose distribution that results in better functional outcome as compared to current treatments based on chemotherapy but especially in less radiation-induced secondary tumors by reducing the irradiated volume of healthy tissues.

\section{Materials and methods}

\subsection{Retinoblastoma collimator}\label{ssec:Collimator}

\begin{figure}
	\begin{center}
		\includegraphics[width=8cm]{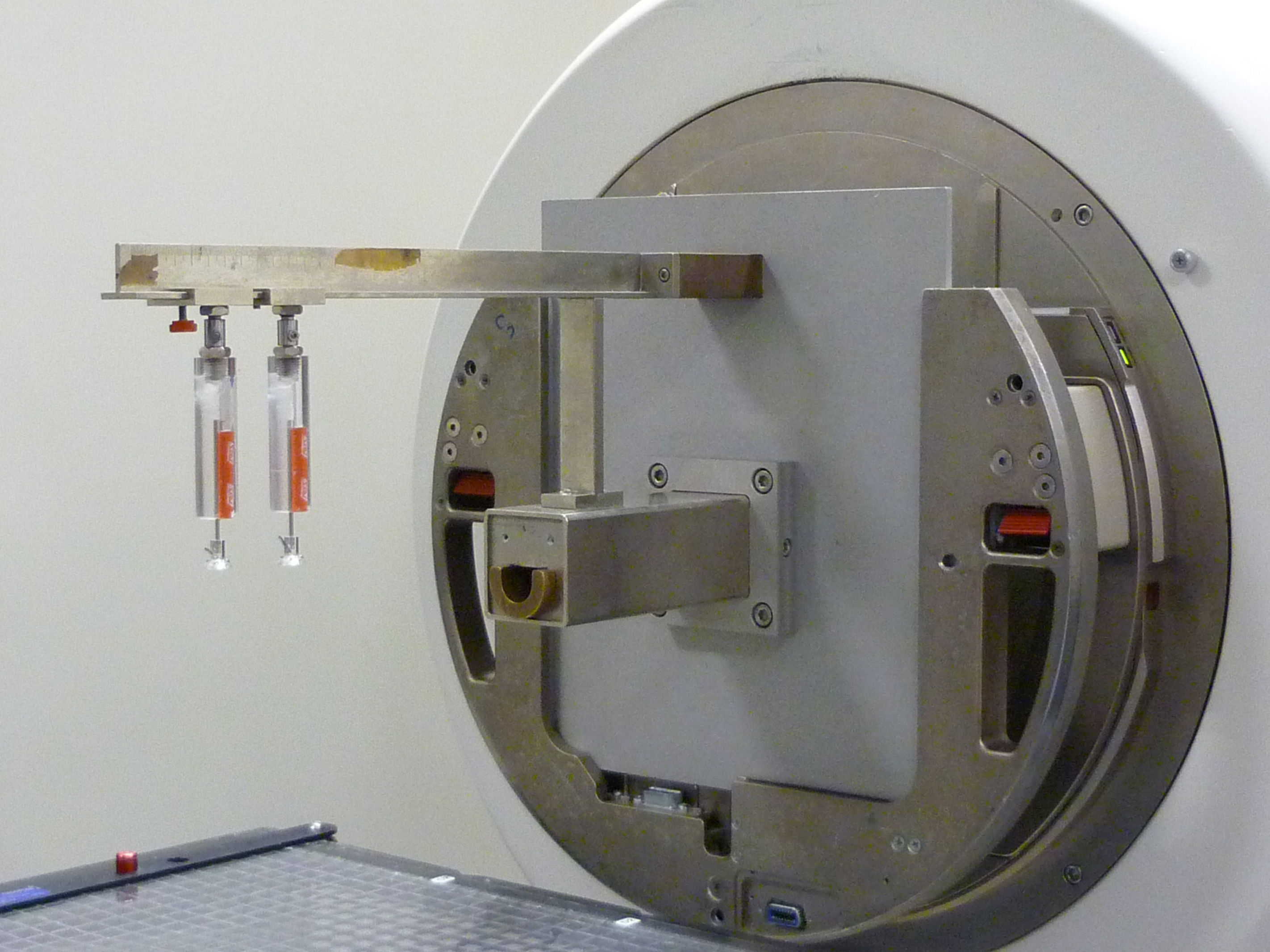} \caption[~]{Photograph of the retinoblastoma collimator, with the optional brass collimator inserted, mounted on a Varian Clinac~2100~C/D. The gantry and collimator angle of the linac are those used for the irradiation of the patient.}
		\label{fig:Picture} 
	\end{center}
\end{figure}

At the University Hospital of Essen a Varian Clinac~2100~C/D operating at 6~MV is used to treat retinoblastoma in patients. A dedicated collimator, inserted in the accessory tray holder, conforms a `D'-shaped off-axis field by means of a long cerrobend tube whose downstream end is located at 17 cm from the isocenter. An optional brass insert can be placed in the dedicated collimator to deliver a smaller field size, which is also `D'-shaped. Thus, patients can be treated either using a larger irradiated area of 5.2~cm$^2$ or a smaller one of 3.1~cm$^2$. Figure~\ref{fig:Picture} shows a picture of the retinoblastoma collimator with the optional brass collimator mounted on a Clinac~2100~C/D. Since the technique is used with pediatric patients who cannot collaborate during the treatment, they must be anesthetized, immobilized and their eyes fixed with vacuum lenses designed to position the eye lenses of the patient with respect to the anterior flat edge of the field. Thus, the collimating device has a fixation system for the eyes and distance scales for determining the eye position with respect to the isocenter and beam central axis. These fixation and positioning devices are attached to the collimator and they can also be seen in the photograph.

\begin{figure}
	\begin{center}
		\includegraphics[height=20.9cm]{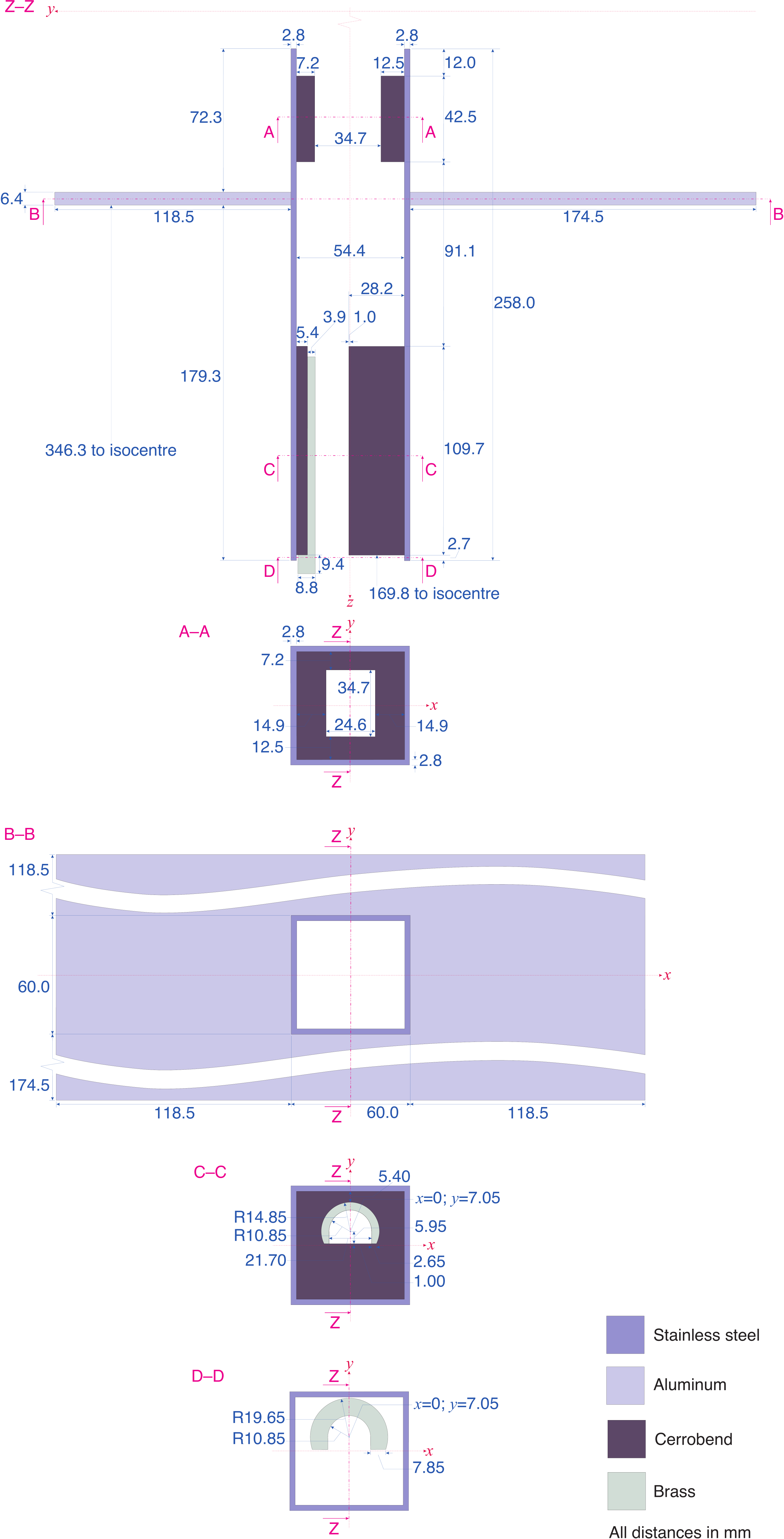} \caption[~]{Blueprints of the dedicated collimator used for retinoblastoma irradiation. The blueprints show the optional brass collimator inserted.} \label{fig:Blueprint} 
	\end{center}
\end{figure}

Figure~\ref{fig:Blueprint} shows the blueprints of the retinoblastoma collimator with the optional brass collimator inserted. Only the parts relevant for conforming the field appear while the fixation and positioning devices are not drawn. The $z$ axis corresponds to the beam central axis with the beam propagating in the direction of increasing $z$. In these blueprints the collimator appears in the position used for irradiating a dosimetry water phantom. The coordinate system plotted in the blueprints is the same used throughout this article. The origin of coordinates is located at the center of the upstream surface of the bremsstrahlung target of the linac, 100~cm away from the isocenter along the $z$-axis. Station B--B in figure~\ref{fig:Blueprint} corresponds to the aluminum plate that is inserted in the accessory tray holder of the linac.

The movable X-jaws are positioned to define a symmetric field about the $y$ axis of width equal to 5.5~cm (each X-jaw is displaced 2.75~cm away from the closed position). In contraposition, the field defined by the Y-jaws is not centered, with the Y-jaws in the negative and positive segments of the $y$ axis displaced  0.7 and 3.5~cm, respectively, from their closed position. Therefore, a $5.5 \times 4.2$~cm$^2$ field is defined with its center at $x=0$~cm and $y=1.4$~cm as measured at 100~cm from the origin of coordinates. The same field defined by the movable jaws is used independently of the presence of the optional brass collimator.

\subsection{Experimental measurements}

Depth dose distributions and lateral profiles for both field sizes were experimentally measured with an IBA water phantom RFA Plus. The gantry rotation angle and the collimator angle were set to zero degrees and the source-to-surface distance equaled 100~cm. With these settings the bar of the positioning and fixation system of the retinoblastoma collimator was partially immersed in the water and, therefore, the movement of the scanning support was restricted on the $y$ direction to avoid collision with the bar. Because neither the reticule center nor the distance indicator are visible while the retinoblastoma collimator is inserted in the Clinac it was necessary to adjust the detector with its effective measurement point placed at the isocenter before mounting the retinoblastoma collimator. However, inserting the retinoblastoma collimator in the accessory tray holder of the Clinac requires lowering the water phantom and rotating the collimator of the linac an angle of 45 degrees. The isocenter level was marked on the water tank before lowering it. All these technical difficulties reduced the position accuracy of the detector. The final position uncertainty of the experimental measurements was estimated to be 0.5~mm.

Relative absorbed dose measurements were performed with a diamond detector with a sensitive diameter of about 2~mm (PTW 60003, Freiburg) which provides good agreement with ionization chambers, and a stereotactic diode with a sensitive diameter of about 0.6~mm (SFD, IBA Dosimetry, Schwarzenbruck) and a better spatial resolution. Experimental measurements were compared to those obtained from a stereotactic ionization pinpoint chamber (PTW 31016, Freiburg). For absolute dose measurements the pinpoint chamber was used with the effective measurement point placed in the center of the field at the maximum depth in the water phantom. A cross calibration was done using reference fields $5 \times 5$~cm$^2$ and $10 \times 10$~cm$^2$ without the retinoblastoma collimator. The actual dose of these fields was determined employing an ionization chamber (PTW M310013, Freiburg). 

The diamond detector was preirradiated with at least 5 Gy. All measurements with the retinoblastoma collimator were done without a reference signal because of the small field. This procedure can be followed because of the high stability dose rate characteristic of the Varian Clinac~2100~C/D.

\subsection{Monte Carlo simulations}

MC simulations were run using the general-purpose MC code system {\sc penelope}~\cite{PENELOPE,Sempau1997,Baro1995} for the coupled transport of electrons, positrons and photons. {\sc penelope} is a set of FORTRAN subroutines that requires a main steering program in charge of defining the primary particle source, tallying the required quantities and applying the necessary variance-reduction techniques; the code {\sc penEasy} was used for that purpose~\cite{Sempau2011}. For estimating the absorbed dose in water it is first necessary to simulate the Varian Clinac~2100~C/D. The geometry, material and configuration files for that linac were automatically generated with the program {\sc penEasyLinac}~\cite{Sempau2011,Brualla2009}. The simulations run with the codes {\sc penelope}/{\sc penEasy}/{\sc penEasyLinac} have been validated in previous publications, in particular for the case of the Varian Clinac~2100~C/D operating at 6~MV, which is the nominal energy used in all experiments, simulations and calculations presented in this article~\cite{Sempau2011,Brualla2009,Panettieri2009,Brualla2009a}. 

{\sc penelope} uses a mixed simulation scheme that classifies electron and positron interactions either as hard or soft events. Hard events, defined as those involving angular deflections or energy losses above certain cutoffs, are simulated in a detailed manner. All soft interactions between two hard events are grouped together and simulated with a single artificial event. The cutoffs are determined by five user-defined transport parameters: C1, controls the average angular deflection between two consecutive hard elastic collisions; C2, is related to the maximum fractional energy loss between two hard elastic collisions; WCC and WCR are the cutoffs for inelastic and bremsstrahlung interactions, respectively. Finally, DSMAX is an upper limit for the step length. The user-defined transport parameters were automatically set by {\sc penEasyLinac} to $\mbox{C1}=\mbox{C2}=0.1$, $\mbox{WCC}=100$~keV, $\mbox{WCR}=20$~keV and the DSMAX value for each constructive element of the linac equaled $1/10$ of its thickness along the beam path. In the simulation all particles are transported until their kinetic energies fall below certain user-defined absorption energies. These energies were set to 100~keV for electrons and positrons, and 20~keV for photons.

The MC simulation for estimating the absorbed dose in a water phantom was divided into three subsequent simulations. First, a simulation was run from the primary electron source downstream to the exit of the gantry, where a phase-space file was tallied. The primary electron source was modeled as a monoenergetic point-like pencil beam with zero divergence and initial kinetic energy equal to 6.26~MeV. The movable jaws were positioned according to the description of the retinoblastoma treatment given in section~\ref{ssec:Collimator}. Secondly, two subsequent simulations used the previously tallied phase-space file as a source and particles were transported downstream of the retinoblastoma collimator with and without the additional brass collimator. These simulations produced two phase-space files at 100~cm from the source, one for the 5.2~cm$^2$ field (without the brass collimator) and another one for the smaller 3.1~cm$^2$ field (with the brass collimator). The geometry of the retinoblastoma collimator was coded using the constructive quadric surface package {\tt pengeom} provided with {\sc penelope} and strictly following the blueprints given in figure~\ref{fig:Blueprint}. Thirdly, the phase-space files tallied downstream of both retinoblastoma collimators were transported each into a water phantom with bin size equal to $0.05 \times 0.05 \times 0.1$~cm$^3$, and two three-dimensional absorbed dose distributions were tallied, one for the 5.2~cm$^2$ field and another one for the 3.1~cm$^2$ field.

\subsection{Calculation with the analytical anisotropic algorithm}

The AAA~\cite{AAA,Ulmer1995,Ulmer2003} as implemented in the treatment planning system Eclipse (Varian) was used to calculate the dose distribution obtained when using the retinoblastoma collimator with its two available field sizes. The versions of the software used were Eclipse 8.9.09 on ARIA 8 with AAA version 8.9.08. With Eclipse it is not possible to simulate the detailed structure of the retinoblastoma collimator shown in figure~\ref{fig:Blueprint} as it was done with {\sc penelope}. The retinoblastoma collimator must be approximated to an aperture block of 10~cm of thickness with transmission of 0.1\% placed on a tray with 100\% transmission. The aperture is shaped according to the field defined by the retinoblastoma collimator at 100~cm from the source. Two apertures were modeled, one for each considered field size. The calculation grid size was chosen to be 0.1~cm whereas the phantom voxel size was $0.086 \times 0.086 \times 0.2$~cm$^3$. The two three-dimensional absorbed dose distributions were exported in DICOM format and then converted into ASCII for comparing with experimental data and MC results.

\section{Results} \label{sec:results}

\begin{figure}[!b]
		\subfigure[]{ \includegraphics[width=.32\columnwidth]{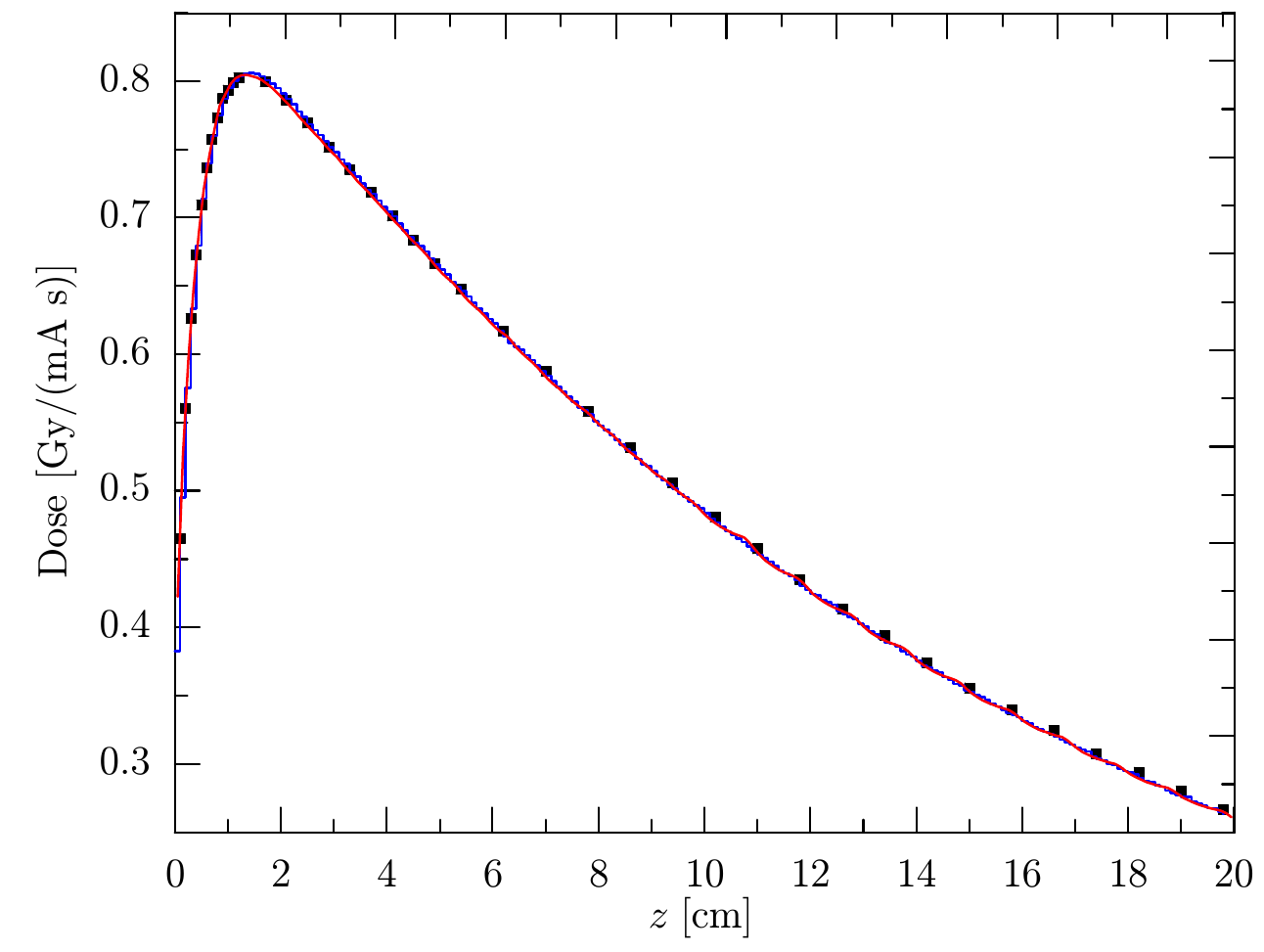}} 
		\subfigure[]{ \includegraphics[width=.32\columnwidth]{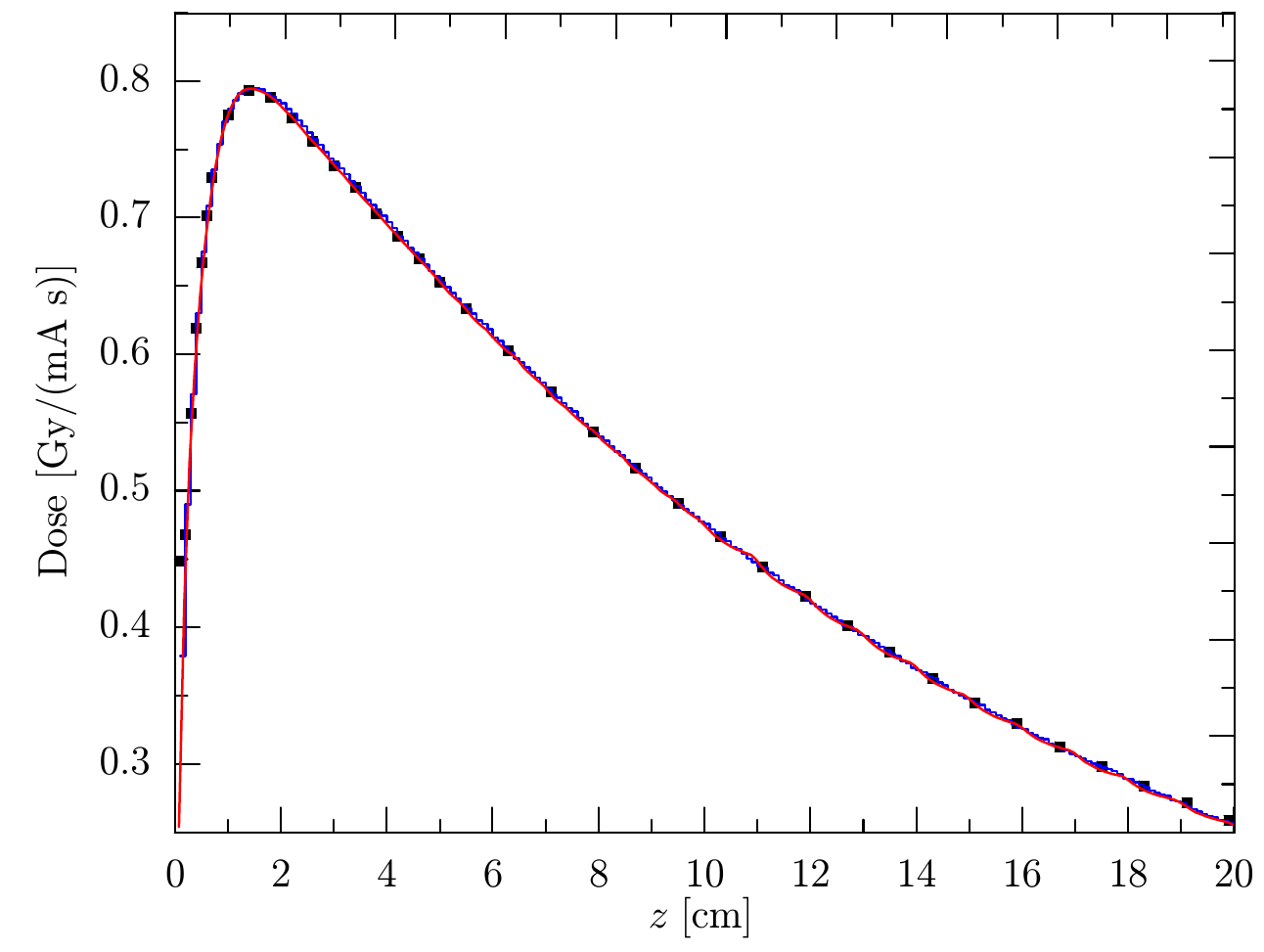}} 
		\caption[~]{ Depth dose distributions at $x=0$~cm and $y=1.3$~cm. Symbols represent experimental data measured with diamond detector, histograms correspond to MC simulations, lines plot AAA results. Results for the 5.2~cm$^2$ (a) and the 3.1~cm$^2$ (b) field sizes. MC statistical uncertainty bars ($1\sigma$) are smaller than symbol size. Experimental data were scaled as explained in the text. For clarity reasons only some experimental points are plotted.} \label{fig:DepthDose} 
\end{figure}

\begin{figure}[!t]
		\subfigure[]{ \includegraphics[width=.32\columnwidth]{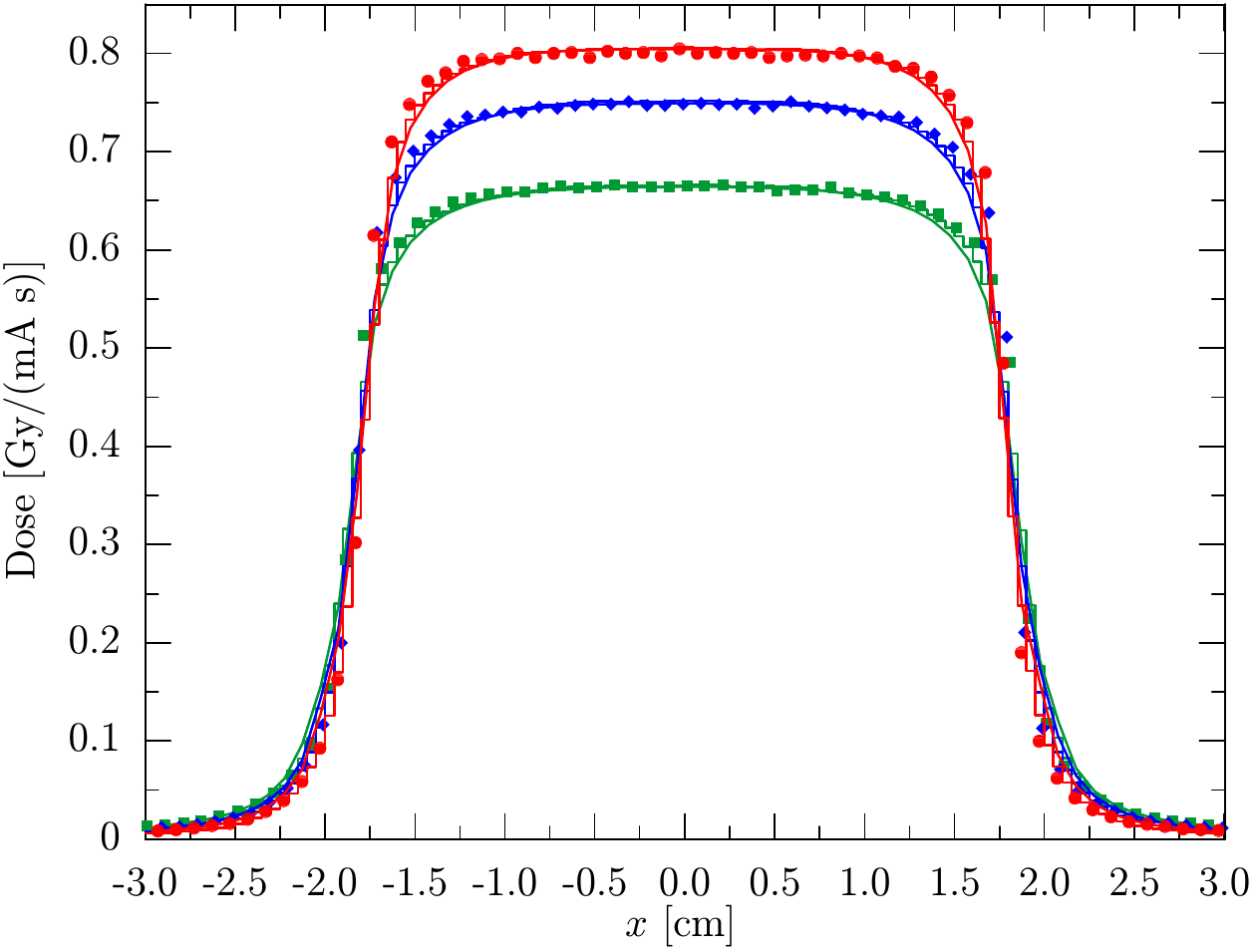}} 
		\subfigure[]{ \includegraphics[width=.32\columnwidth]{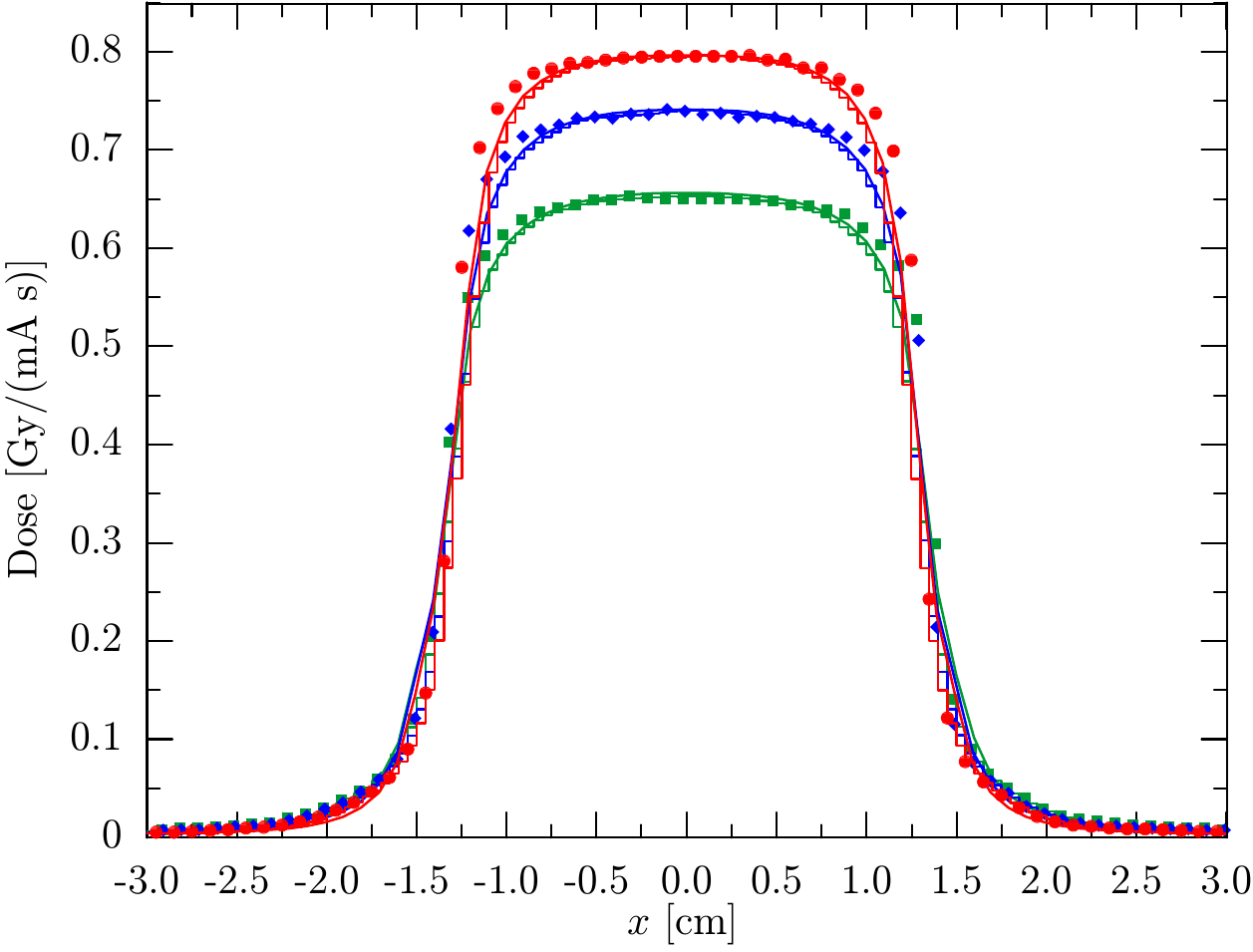}} 
		\caption[~]{ Lateral profiles along the $x$ axis for $y=1.3$~cm plotted at depths $z=1.5$ (red circles), 3.0 (blue diamonds) and 5.0~cm (green squares). Symbols represent experimental data measured with SFD diode, histograms correspond to MC simulations, lines plot AAA results. Results for the 5.2~cm$^2$ (a) and the 3.1~cm$^2$ (b) field sizes. Other details are the same as in figure~\ref{fig:DepthDose}.} \label{fig:LateralX} 
\end{figure}

\begin{figure}[!th]
		\subfigure[]{ \includegraphics[width=.32\columnwidth]{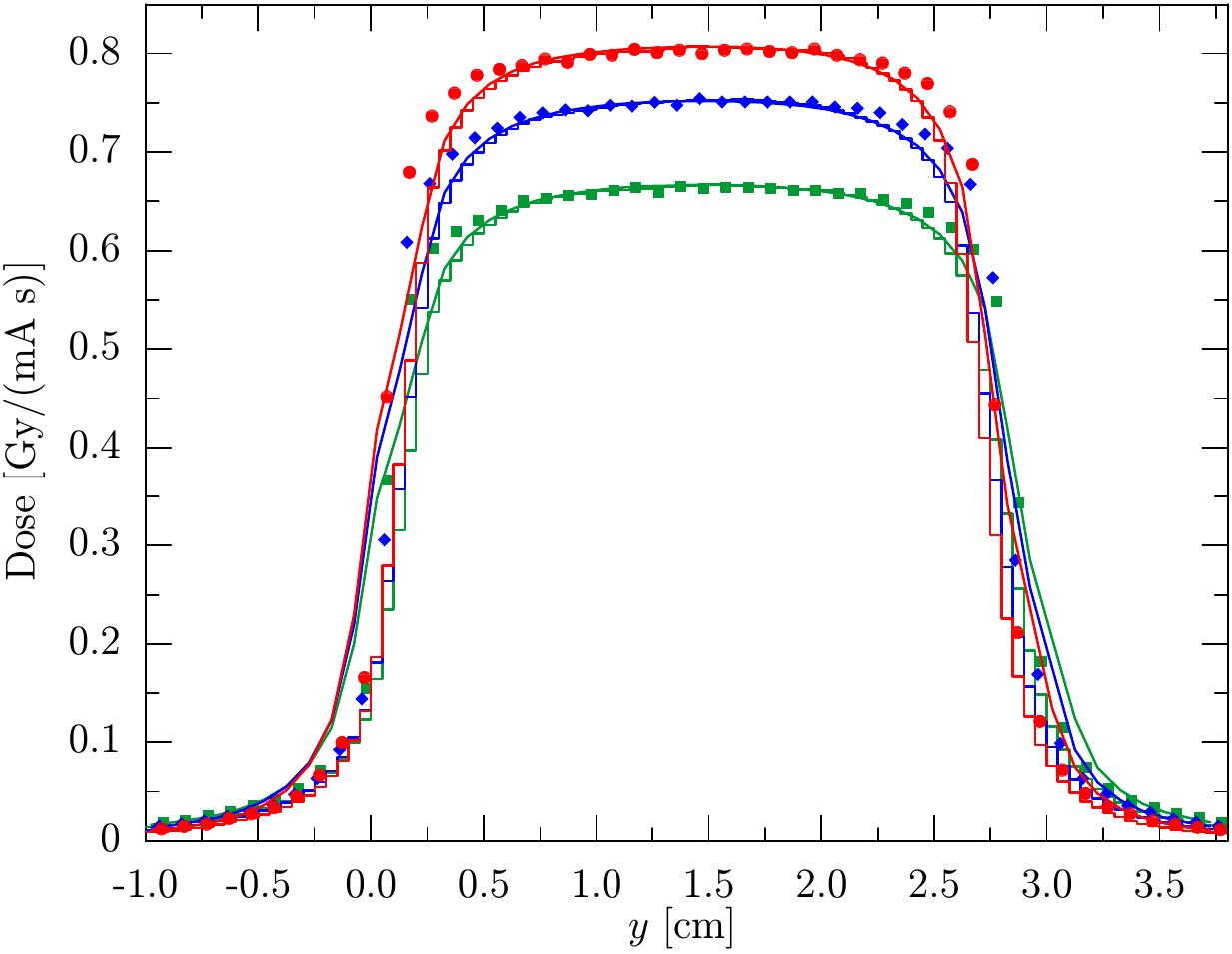}} 
		\subfigure[]{ \includegraphics[width=.32\columnwidth]{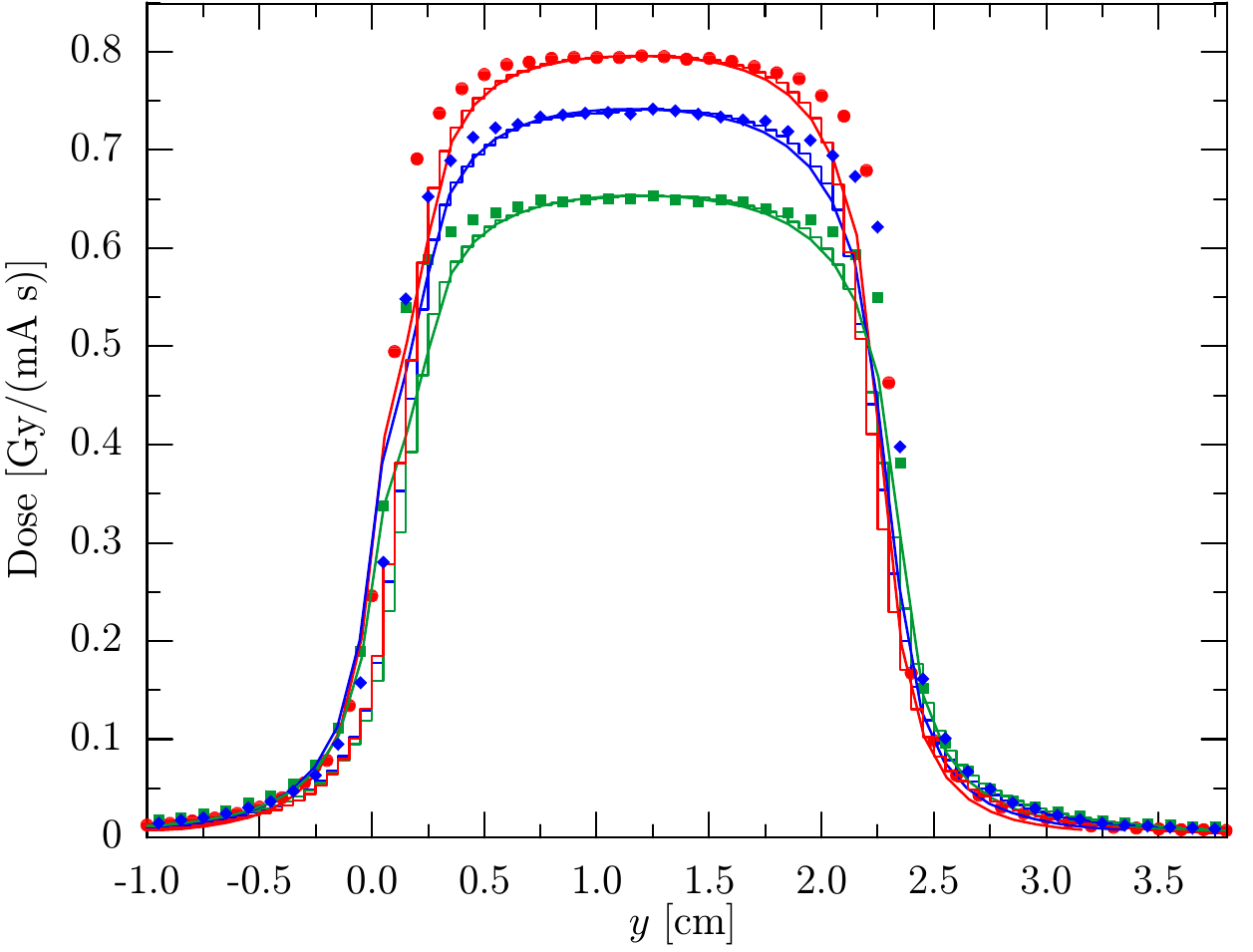}} 
		\caption[~]{ Lateral profiles along the $y$ axis for $x=0$~cm plotted at depths $z=1.5$ (red circles), 3.0 (blue diamonds) and 5.0~cm (green squares). Symbols represent experimental data measured with SFD diode, histograms correspond to MC simulations, lines plot AAA results. Results for the 5.2~cm$^2$ (a) and the 3.1~cm$^2$ (b) field sizes. Other details are the same as in figure~\ref{fig:DepthDose}.} \label{fig:LateralY} 
\end{figure}

Depth dose distributions and lateral profiles obtained for both field sizes, namely 5.2 and 3.1~cm$^2$, are plotted in figures~\ref{fig:DepthDose}, \ref{fig:LateralX} and \ref{fig:LateralY}. In all plots symbols were used for experimental data, histograms for MC results and lines for the AAA calculations. Doses per primary particle are expressed in $\mbox{Gy}/(\mbox{mA} \, \mbox{s})$ as computed from {\sc penelope}, whence the dose in Gy can be calculated knowing the current intensity at the target in mA and the irradiation time in~s. The experimental relative dose profiles were scaled to match the maximum absorbed dose obtained with {\sc penelope} using the same scaling factor for all curves for a given field size. The same procedure was applied to scale the data obtained with the AAA. 

The gamma test~\cite{Low1998} was used for a quantitative comparison of the plotted dose profiles. Table~\ref{tab:GammaTest} shows the comparison of the results obtained with {\sc penelope} and the AAA against the experimental data used as reference data set. All absorbed dose profiles were compared using the following three criteria: distance-to-agreement (DTA) of 0.1~cm and a discrepancy of 1\% of the dose, 0.2~cm and 2\%, and 0.3~cm and 3\%. Additionally, for the comparison of AAA results with respect to the experimental data the criteria 0.3~cm and 6\% was also used. The values reported on table~\ref{tab:GammaTest} are the percentage of voxels that exceed a gamma index equal to 1, that is, the percentage of voxels that fail the test, for each dose profile, criteria and computing algorithm. 

\begin{table}
\caption{Gamma test for comparing {\sc penelope} results against the experimental data and AAA results against the experimental data. See text for details.} \label{tab:GammaTest} 
\begin{center}%
\begin{tabular}{cccccccccc}%
\hline \hline
 field                       & profile           &  \multicolumn{3}{c}{{\sc penelope}}    &~~&  \multicolumn{4}{c}{AAA} \\ \cline{3-5}\cline{7-10}
                             &                   & 0.1 cm, 1\% & 0.2 cm, 2\% & 0.3 cm, 3\% && 0.1 cm, 1\% & 0.2 cm, 2\% & 0.3 cm, 3\%& 0.3 cm, 6\% \\ \hline
 \multirow{7}{*}{5.2 cm$^2$} & depth dose        & 3           & 2           & 0           && 7           & 2           & 0          & 0           \\
                             & $x$ at $z=1.5$ cm & 7           & 6           & 2           && 34          & 16          & 10         & 0           \\ 
                             & $y$ at $z=1.5$ cm & 30          & 11          & 3           && 39          & 22          & 6          & 3           \\ 
                             & $x$ at $z=3.0$ cm & 11          & 4           & 0           && 27          & 13          & 5          & 2           \\ 
                             & $y$ at $z=3.0$ cm & 21          & 5           & 2           && 35          & 19          & 8          & 3           \\ 
                             & $x$ at $z=5.0$ cm & 12          & 6           & 0           && 15          & 8           & 6          & 0           \\ 
                             & $y$ at $z=5.0$ cm & 21          & 5           & 0           && 35          & 19          & 7          & 2           \\ \hline
 \multirow{7}{*}{3.1 cm$^2$} & depth dose        & 2           & 1           & 0           && 5           & 2           & 0          & 0           \\
                             & $x$ at $z=1.5$ cm & 22          & 8           & 3           && 33          & 11          & 3          & 0           \\ 
                             & $y$ at $z=1.5$ cm & 25          & 9           & 3           && 53          & 34          & 18         & 3           \\ 
                             & $x$ at $z=3.0$ cm & 29          & 5           & 0           && 40          & 19          & 9          & 2           \\ 
                             & $y$ at $z=3.0$ cm & 31          & 12          & 3           && 43          & 25          & 10         & 0           \\ 
                             & $x$ at $z=5.0$ cm & 23          & 5           & 0           && 46          & 18          & 7          & 0           \\ 
                             & $y$ at $z=5.0$ cm & 31          & 8           & 0           && 45          & 20          & 12         & 0           \\ \hline\hline
\end{tabular}                                                                                                                       
\end{center}
\end{table}

\section{Discussion and conclusions} \label{sec:conclusions}

For both fields there is an excellent agreement between the depth dose distributions obtained with MC simulations and the experimental data. Regarding the AAA results on depth dose distributions, the agreement with the experimental data is nearly as good as that obtained with MC simulations. However, a close inspection of the depth dose curves obtained with the AAA reveals a sinusoidal pattern on the curve for depths greater than 10~cm. This artifact already appears when the kernel beamlets of the AAA are generated and it is observed in depth dose calculations of reference fields smaller than $3 \times 3$~cm$^2$ in the beam configuration.

The agreement between simulated results and experimental data in relation to lateral profiles is substantially worse with respect to the degree of agreement observed for the depth dose distributions (table~\ref{tab:GammaTest}). {\sc penelope} simulations agree reasonably well with the experimental data with a DTA less than 0.3~cm and a dose difference below 3\%. The largest differences between these two data sets can be ascribed on the shoulder region of the lateral profiles. When comparing results from the AAA with experimental data a zero percent of voxels exceeding gamma index equal to one is reached with the criteria 0.3~cm and 6\%. In the case of the AAA differences appear not only on the shoulders but also on the penumbra tails.

Due to the off-axis character of the irradiated fields, the penumbra tails calculated with the AAA along the $y$ axis show marked discrepancies with respect to the experimental data, while similar discrepancies do not appear in the profiles computed along the $x$ axis (figures~\ref{fig:LateralX}, \ref{fig:LateralY} and table~\ref{tab:GammaTest}). Additionally, for both considered fields an artifact, a `kink' in the lateral profile, appears on the AAA data at $y=0$~cm (figure~\ref{fig:LateralY}). The reason for this kink seems to be the following: the evaluated version of AAA uses for its calculations an input fluence map whose pixel size is fixed to 2.5~mm$^2$ with a gridding in which the point $x=y=0$ coincides with the interface among four pixels. As it can be seen from the blueprints, the irradiated field is 1~mm away from the central beam axis, in other words, the central beam axis is obscured by a cerrobend block. Therefore, the fluence pixel from $y=0$ to $y=2.5$~mm corresponds to a region with transmission coefficient of 0.1\% along its first millimeter and a transmission coefficient of 100\% along the last 1.5~mm. A test calculation was performed in which the flat side of the `D'--shaped field was positioned on the central beam axis and the kink angle was much reduced and displaced to a higher dose region of the profile, almost disappearing.

The kink in the lateral profiles computed with the AAA appears in a delicate position from the therapeutical point of view. The fixation system of the retinoblastoma collimator aligns the main symmetry axis of the eye lens with the $y$ axis. The eye lens is positioned such that its posterior pole coincides with $x=y=0$~cm. Therefore, the whole eye lens is located at $y<0$~cm, that is, protected by the cerrobend. Owing to the kink the width of the penumbra tail of the AAA results is overestimated by nearly 2~mm, more than half of the thickness of the eye lens, therefore overestimating the actual dose delivered to it. 

Previous studies on 6~MV beams have shown that the discrepancy of the doses obtained in water using the AAA and experimental data is less than 2\% of the maximum absorbed dose~\cite{VanEsch2006,Ding2007,Fogliata2006,Panettieri2009}. However, in these studies the smallest considered field is $3 \times 3$~cm$^2$ centered. In contradistinction, the smallest field considered in the present study is 3.1~cm$^2$, highly conformed and off-axis. Under these more difficult conditions the discrepancy between the AAA and the experiment reaches 0.3~cm and 6\%. Although these discrepancies are relevant from the therapeutical point of view, it is still possible to consider the AAA for routine treatment planning of retinoblastoma patients, provided the limitations of the algorithm are known and taken into account by the medical physicist. Knowledge on the limitations of the algorithm can be achieved by a combination of experiments and MC simulations. Future research will be aimed at improving the design of the retinoblastoma collimator and the irradiation technique. For this purpose MC simulations are indispensable.

\begin{acknowledgments}
	
LB gratefully acknowledges financial support from the Deutsche Forschungsgemeinschaft project BR~4043/1-1. PAM is grateful to the International Atomic Energy Agency for financial support through a Ph.D. fellowship (project code RLA/06/061). AML acknowledges partial financial support from the Spanish Ministerio de Ciencia e Innovaci\'on (project no.\ FPA2009-14091-C02-02) and the Junta de Andaluc\'{\i}a (project no.\ FQM-0220). JS is grateful to the Spanish Ministerio de Ciencia e Innovaci\'on (project no.\ FPA2009-14091-C02-01) and to the Spanish Networking Research Center CIBER-BBN for partial financial support. 
 
\end{acknowledgments}


\begin{thebibliography}{99}
	
	\bibitem{Friend1986} S.H. Friend, R.A. Bernards, S. Rogelj et al., ``A human DNA segment having properties of the gene that predisposes to retinoblastoma and osteosarcoma,'' Nature {\bf 323}, 643--646 (1986). 
	
	\bibitem{Sauerwein1997} W. Sauerwein, W. H\"opping and N. Bornfeld, ``Radiotherapy for retinoblastoma. Treatment strategies,'' Front. Radiat. Ther. Oncol. {\bf 30}, 93--96 (1997).
	
	\bibitem{Abramson1998} D.H. Abramson and C.M. Frank, ``Second nonocular tumors in survivors of bilateral retinoblastoma: a possible age effect on radiation-related risk,'' Ophthalmology {\bf 105}, 573--580 (1998).
	
	\bibitem{Dommering2011} C.J. Dommering, T. Marees, A.H. Hout et al., ``RB1 mutations and second primary malignancies after hereditary retinoblastoma,'' Fam. Cancer doi:10.1007/s10689-011-9505-3 (2011).
	
	\bibitem{Turaka2011} K. Turaka, C.L. Shields, A.T. Meadows, A. Leahey, ``Second malignant neoplasms following chemoreduction with carboplatin, etoposide, and vincristine in 245 patients with intraocular retinoblastoma,'' Pediatr. Blood Cancer doi:10.1002/pbc.23278 (2011).
	
	\bibitem{Vasudevan2010} V. Vasudevan , M.C. Cheung, R. Yang, Y. Zhuge, L.G. Koniaris and J.E. Sola, ``Pediatric solid tumors and second malignancies: characteristics and survival outcomes,'' J. Surg. Res. {\bf 160}, 184--189 (2010).
	
	\bibitem{Shields1996} C.L. Shields, P. De Potter, B.P. Himelstein, J.A. Shields, A.T. Meadows, J.M. Maris, ``Chemoreduction in the initial management of intraocular retinoblastoma,'' Arch. Ophthalmol. {\bf 114}, 1330--1338 (1996).
	
	\bibitem{Gallie1996} B.L. Gallie, A. Budning, G. DeBoer, et al., ``Chemotherapy with focal therapy can cure intraocular retinoblastoma without radiotherapy,'' Arch. Ophthalmol. {\bf 114}, 1321--1328 (1996).
	
	\bibitem{Gobin2011} Y.P. Gobin, I.J. Dunkel B.P. Marr, S.E. Brodie and D.H. Abramson, ``Intra-arterial chemotherapy for the management of retinoblastoma: four-year experience,'' Arch. Ophthalmol. {\bf 129}, 732--737 (2011).
	
	\bibitem{Schueler2006a} A.O. Schueler, D. Fl{\"u}hs, G. Anastassiou, C. Jurklies, W. Sauerwein and N. Bornfeld, ``Beta-ray brachytherapy of retinoblastoma: feasibility of a new small-sized ruthenium-106 plaque,'' Ophthalmic. Res. {\bf 38}, 8--12 (2006).
	
	\bibitem{Schueler2006b} A.O. Schueler, D. Fl{\"u}hs, G. Anastassiou et al., ``$\beta$-ray brachytherapy with $^{106}$Ru plaques for retinoblastoma,'' Int. J. Radiat. Oncol. Biol. Phys. {\bf 65}, 1212--1221 (2006).
	
	\bibitem{Munier2008} F.L. Munier, J. Verwey, A. Pica, et al.``New developments in external beam radiotherapy for retinoblastoma: from lens to normal tissue-sparing techniques,'' Clin. Experiment. Ophthalmol. {\bf 36}, 78--79 (2008).
	
	\bibitem{Qaddoumi2012} I. Qaddoumi, J.K. Bass, J. Wu et al., ``Carboplatin-associated ototoxicity in children with retinoblastoma,'' J. Clin. Oncol. doi:10.1200/JCO.2011.36.9744 (2012).
	
	\bibitem{Gombos2007} D.S. Gombos and P. Chevez-Barrios, ``Current treatment and management of retinoblastoma,'' Curr. Oncol. Rep. {\bf 9}, 453--458 (2007).
	
	\bibitem{Felix2001} C.A. Felix, ``Leukemias related to treatment with DNA topoisomerase II inhibitors,'' Med. Pediatr. Oncol. {\bf 36}, 525--535 (2001).
	
	\bibitem{Schipper1983} J. Schipper, ``An accurate and simple method for megavoltage radiation therapy of retinoblastoma,'' Radiother. Oncol. {\bf 1}, 31--41 (1983).
	
	\bibitem{Schipper1997} J. Schipper, S.M. Imhoff and K.E.W.P. Tan, ``Precision megavoltage external beam radiation therapy for retinoblastoma,'' Front. Radiat. Ther. Oncol. {\bf 30}, 65--80 (1997).
	
	\bibitem{Sauerwein2009} W. Sauerwein and C.E. Stannard, ``Auge und Orbita,'' in \emph{Radioonkologie Band 2: Klinik}, edited by M. Bamberg, M. Molls and H. Sack (W. Zuckschwerdt Verlag, M\"unchen Wien New York, 2009), pp. 294--316.
	
	\bibitem{AAA} J. Sievinen, W. Ulmer and W. Kaissl, ``AAA photon dose calculation model in Eclipse,'' Varian Medical Systems, 2006.

	\bibitem{Ulmer1995} W. Ulmer W and D. Harder, ``A triple gaussian pencil beam model for photon beam treatment planning'' Z. Med. Phys. {\bf 5} 25--30 (1995).

	\bibitem{Ulmer2003} W. Ulmer and W. Kaissl, ``The inverse problem of a Gaussian convolution and its application to the finite size of the measurement chambers/detectors in photon and proton dosimetry,'' Phys. Med. Biol. {\bf 48}, 707--727 (2003).
	
	\bibitem{PENELOPE} F. Salvat, J.M. Fern\'andez-Varea and J. Sempau, \emph{PENELOPE 2008---A Code System for Monte Carlo Simulation of Electron and Photon Transport}, (OECD Nuclear Energy Agency, Issy-les-Moulineaux, France, 2009.)
	
	\bibitem{Sempau1997} J. Sempau, E. Acosta, J. Bar\'o, J.M. Fern\'andez-Varea and F. Salvat, ``An algorithm for Monte Carlo simulation of coupled electron-photon transport,'' Nucl. Instrum. Meth. B {\bf 132}, 377--390 (1997).
	
	\bibitem{Baro1995} J. Bar\'o, J. Sempau, J.M. Fern\'andez-Varea and F. Salvat, ``{\sc penelope}: An algorithm for Monte Carlo simulation of the penetration and energy loss of electrons and positrons in matter,'' Nucl. Instrum. Meth. B {\bf 100}, 31--46 (1995).
	
	\bibitem{Sempau2011} J. Sempau, A. Badal and L. Brualla, ``A {\sc penelope}-based system for the automated Monte Carlo simulation of clinacs and voxelized geometries---application to far-from-the-axis fields,'' Med. Phys. {\bf 38}, 5887--5895 (2011).
	
	\bibitem{Brualla2009} L. Brualla, F. Salvat and R. Palanco-Zamora, ``Efficient Monte Carlo simulation of  multileaf collimators using geometry-related variance-reduction techniques,'' Phys. Med. Biol. {\bf 54}, 4131--4149 (2009).

	\bibitem{Panettieri2009} V. Panettieri, P. Barsoum, M. Westermark, L. Brualla and I. Lax, ``AAA and PBC calculation accuracy in the surface build-up region in tangential beam treatments. Phantom and breast case study with the Monte Carlo code {\sc penelope},'' Radiother. Oncol. {\bf 93}, 94--101 (2009).
	
	\bibitem{Brualla2009a} L. Brualla, R. Palanco-Zamora, A. Wittig, J. Sempau and W, Sauerwein, ``Comparison between penelope and eMC in small electron fields,'' Phys. Med. Biol. {\bf 54}, 5469--5481 (2009).

\bibitem{Low1998} D. Low, W. Harms, S. Mutic and J. Purdy, ``A technique for the quantitative evaluation of dose distributions,'' Med. Phys. {\bf 25}, 656--661 (1998).	
	
	\bibitem{VanEsch2006} A. Van Esch, L. Tillikainen, J. Pyykkonen et al. ``Testing of the analytical anisotropic algorithm for photon dose calculation,'' Med. Phys. {\bf 33}, 4130--4148, (2006).
	
	\bibitem{Ding2007} G.X. Ding, D.M. Duggan and C.W. Coffey, ``Comment on `Testing of the analytical anisotropic algorithm for photon dose calculation' [Med. Phys. {\bf 33}, 4130--4148 (2006)],'' Med. Phys. {\bf 34}, 3414 (2007).
		
	\bibitem{Fogliata2006} A. Fogliata, G. Nicolini, E. Vanetti, A. Clivio and L. Cozzi, ``Dosimetric validation of the anisotropic analytical algorithm for photon dose calculation: fundamental characterization in water,'' Phys. Med. Biol. {\bf 51}, 1421--1438 (2006).
	
\end{thebibliography}
\end{document}